\documentclass[12pt]{article}

\usepackage{amsmath,amssymb,a4wide}
\usepackage{epsfig}

\title{
\bf Area distribution of the \\
planar random loop boundary
}

\author{\sc Christoph~Richard\\
\\
Institut f\"ur Mathematik, Universit\"at Greifswald\\
Jahnstr.~15a, 17487 Greifswald, Germany}

\begin{document}

\maketitle

\begin{abstract}
We numerically investigate the area statistics of the outer boundary
of planar random loops, on the square and triangular
lattices.  Our Monte Carlo simulations suggest that the underlying
limit distribution is the Airy distribution, which was recently found to
appear also as the area distribution in the model of self-avoiding loops.
\end{abstract}

\section{Introduction}

In recent years, a number of predictions for critical exponents
related to random planar curves have been proved using probabilistic
methods.  Particular examples are the proof of old conjectures
concerning exponents of critical percolation on the triangular lattice
\cite{SW01}, and the proof of the conjecture that the Hausdorff
dimension of the outer boundary of planar Brownian motion is $4/3$
\cite{LSW01}.  The latter conjecture was stated in 1982 on the basis
of an investigation of random loops on the square lattice \cite{M82},
while a non-rigorous derivation using methods of quantum gravity was
given in 1998 \cite{Dup98}.  

A main ingredient of the approach \cite{SW01, LSW01} is a family
$SLE_{\kappa}$ of conformally invariant stochastic processes, called
Schramm-Loewner evolution (sometimes also stochastic Loewner evolution).
Critical percolation and the outer boundary of planar Brownian motion are
described by $SLE_{6}$.  More generally, it is conjectured that
Schramm-Loewner evolution describes critical Fortuin-Kasteleyn clusters
\cite{W03}.  The area distribution of such clusters has been analysed recently
\cite{CZ02}.

Schramm-Loewner evolution is also connected to the self-avoiding walk model
\cite{MS93}.  It has recently been shown that, if the continuum limit of
self-avoiding walks exists and is conformally invariant, it
is described by $SLE_{8/3}$ \cite{LSW02}.  This led to another
derivation of the critical exponents of the self-avoiding walk
$\nu=3/4$ and $\gamma=43/32$, and to predictions of the distribution
of some random variables related to the self-avoiding walk problem.  Some of
these have been numerically tested recently \cite{K01,K02}.  Moreover, there
are predictions for the self-avoiding polygon model \cite{MS93} based on
$SLE_{8/3}$.  Essentially, the scaling limit of self-avoiding polygons is
predicted to be described by the union of two Brownian excursions
\cite{LSW02}.  Also, it is predicted that the scaling limit coincides with the
outer boundary of loops of Brownian motion \cite{LSW02}.  Since the area under
a Brownian excursion is Airy distributed \cite{L84}, one thus expects
the Airy distribution\footnote{The Airy distribution appears
in a variety of other contexts such the path lengths in trees and the
analysis of the linear probing hashing algorithm, see \cite{FL01} for a review.} 
as the area distribution of the outer boundary of
Brownian motion. 

A completely different, independent approach to the self-avoiding polygon
problem is inspired by the analysis of exactly solvable models of lattice
polygons, counted by perimeter and area.  A conjecture for the scaling
function of self-avoiding polygons \cite{RGJ01,C01} has recently been tested
\cite{RGJ01,J03,RJG03}, on the basis of an extrapolation of exact enumeration
data and Monte Carlo simulations on the square, triangular and hexagonal
lattices.  The validity of the conjecture implies that the asymptotic
distribution of area of self-avoiding polygons is the Airy distribution.

Within this framework, the conjecture for the scaling function of self-avoiding
polygons arose from an analysis of polygon models whose perimeter and
area generating function satisfy a $q$-algebraic functional equation
\cite{RGJ01}, compare also \cite{D99}.  It was shown that if their perimeter
generating function displays a square-root singularity as the dominant
singularity, then the mean area $a_m$ of a polygon of perimeter $m$
generically grows as $a_m\sim A m^{2\nu}$ as $m\to\infty$, where $\nu=3/4$.
This implies a dimension $1/\nu=4/3$ for such polygons.  Moreover, a certain
scaling function of Airy-type generically arises \cite{R02}.  Since the
prediction of the dimension of self-avoiding polygons $4/3$ was also supported
by conformal field theory arguments and numerical analysis \cite{MS93}, it was
natural to ask whether self-avoiding polygons might have the same scaling
function.  Noting that the outer boundary of planar random loops also has 
dimension $4/3$, one might suspect that the Airy distribution appears here as 
well.

In this article, we test the conjecture that the Airy distribution
arises as the limit distribution of the outer boundary of planar random
loops, based on Monte Carlo simulations on the square and triangular
lattices.  Our conjecture arises from the connection to the self-avoiding
polygon model, which may be viewed from either the stochastic perspective or
from the exactly solvable polygon model perspective, as indicated above.  We
find the conjecture of an Airy distribution for the limit distribution of the
outer boundary of planar random loops satisfied, within numerical accuracy.

The article is organised as follows.  In the following section, we will
gather basic results about the Airy distribution.  The third
section explains our algorithms and presents the analysis of data. 
We then summarise our results and discuss possible future work.  In the 
appendix, we discuss the connection between scaling functions and limit
distributions. This motivates the particular form of our prediction for the 
area moments in section two.

\section{Airy distribution}

A random variable $X$ is said to be Airy distributed%
\footnote{
For the definition of the Airy distribution and its properties we
follow \cite{FL01}.}
if its moments
are given by
\begin{equation}\label{form:Airydistr}
\mathbb E[X^k] = -\frac{\Gamma (-1/2)}{\Gamma((3k-1)/2)} \Omega_k,
\end{equation}
where the {\it Airy constants} $\Omega_k$ are
determined by the quadratic recurrence
\begin{equation}\label{form:om}
\Omega_0=-1, \qquad 2 \Omega_k=(3k-4)k\Omega_{k-1} + \sum_{j=1}^{k-1}
\binom{k}{j} \Omega_j\Omega_{k-j} \qquad (k\ge 1).
\end{equation}
The first few numbers $\Omega_k$ are
$-1$, $1/2$, $5/4$, $45/4$, $3315/16$, $25425/4$, $18635625/64$, $18592875$.
The moments uniquely define a probability distribution, which is
called the Airy distribution.  The name relates to the fact
that the numbers $\Omega_k$ also appear in the asymptotic expansion
\begin{equation}\label{form:logAi}
\frac{\rm d}{{\rm d}s} \log \mbox{Ai}(s) \sim \sum_{k=0}^\infty
\frac{(-1)^k}{k!}\frac{\Omega_k}{2^k}
\frac{1}{s^{(3k-1)/2}} \qquad (s\to \infty),
\end{equation}
where $\mbox{Ai}(z)=\frac{1}{\pi}\int_0^\infty \cos(t^3/3+tz)$ is the
Airy function.

In the following section, we will analyse the conjecture that the outer
boundary of planar random loops is Airy distributed, using a Monte
Carlo simulation, both for the square and triangular lattices.  More
precisely, denote by ${\widetilde p}_{m,n}$ the number of sampled
outer boundaries of planar random loops with perimeter $m$ and area
$n$.  (An algorithmic definition of the outer boundary of a random loop will be
presented in the following section.) Define a corresponding random variable
$Y_m$ via
\begin{equation}
\mathbb P(Y_m=n)=\frac{{\widetilde p}_{m,n}}{\sum_n {\widetilde p} _{m,n}}.
\end{equation}
We will numerically analyse the area moments $\mathbb E[Y_m^k]$ under
the assumption that
\begin{equation}\label{form:val}
\mathbb E[Y_m^k] = \frac{\sum_n n^k {\widetilde p}_{m,n}}{\sum_n {\widetilde
    p}_{m,n}}\sim -\frac{\Gamma(-1/2)}{\Gamma((3k-1)/2)} \left( C
    m^{3/2}\right)^k \Omega_k \qquad (m\to\infty)
\end{equation}
for some constant $C>0$.  In comparison to (\ref{form:Airydistr}), the factor
$m^{3k/2}$ accounts for the growth of the area moments in powers of $m^{3/2}$.  
This formula correctly describes the area statistics of the self-avoiding 
polygon model on the square and triangular lattice, with ${\widetilde p}_{m,n}$ 
the number of (sampled) self-avoiding polygons of perimeter $m$ and area $n$. 
(The perimeter of a polygon is the number of edges, its area is the number of 
enclosed squares or triangles.)  See the appendix for a derivation of 
(\ref{form:val}) from the conjectured scaling function of self-avoiding 
polygons.
 
\section{Numerical analysis}

\begin{figure}[htb]
\begin{center}
\hfill
\begin{minipage}[c]{0.45\textwidth}
\center{\epsfig{file=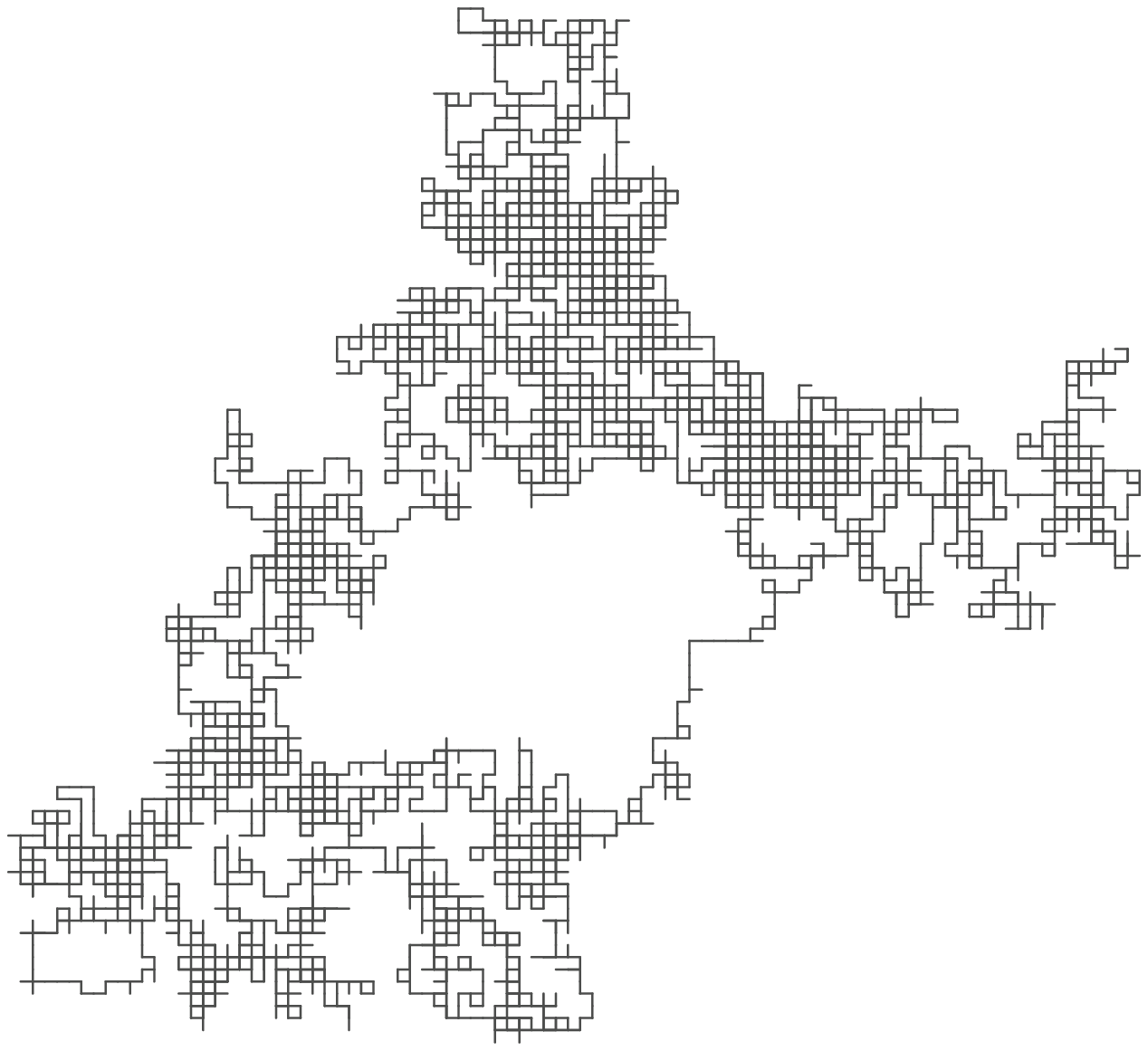,width=6cm}}
\end{minipage}
\hfill
\begin{minipage}[c]{0.45\textwidth}
\center{\epsfig{file=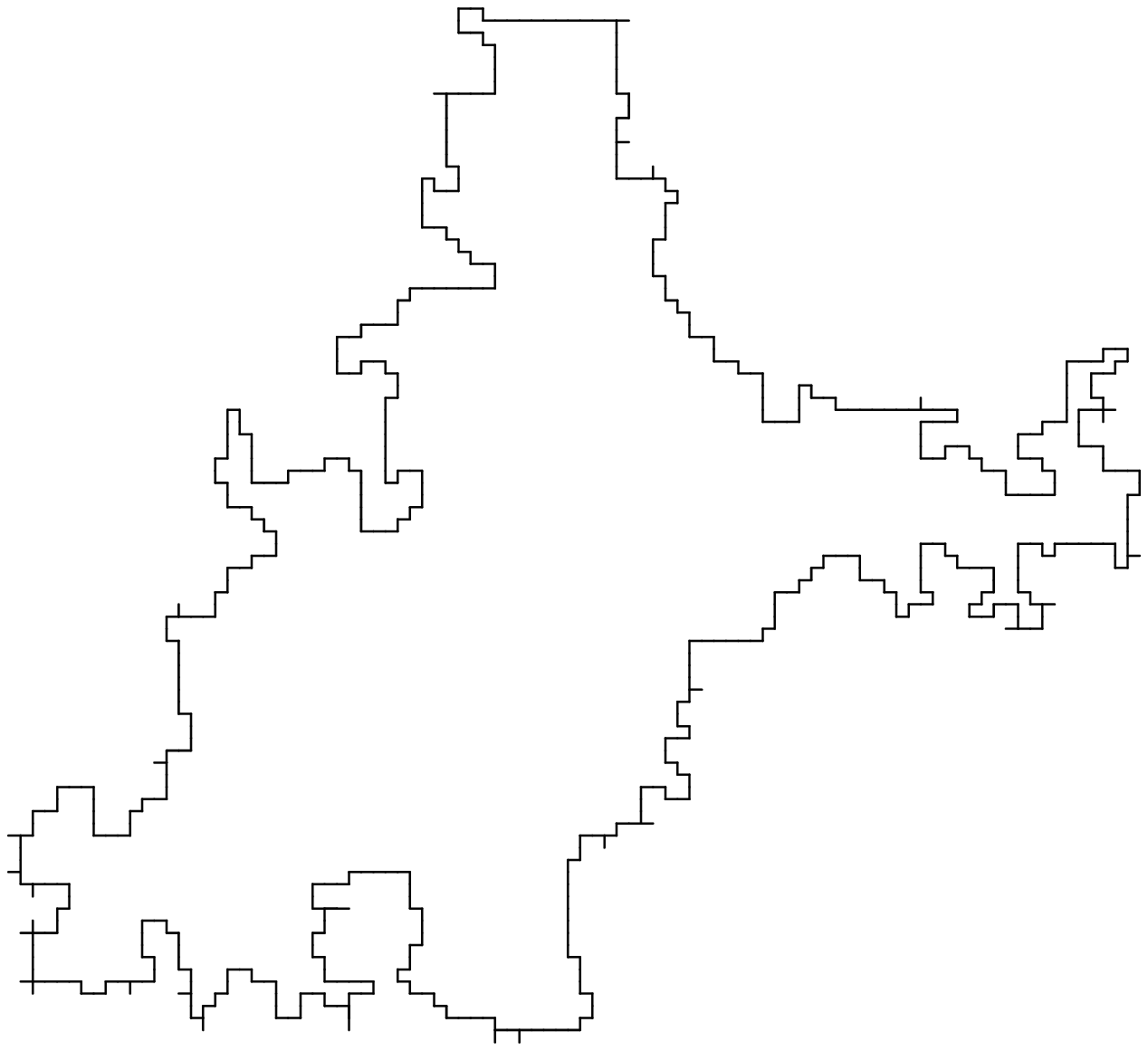,width=6cm}}
\end{minipage}
\hfill
\end{center}
\caption{\label{fig:snapshot}
\small A square lattice random loop (left) and the corresponding outer
boundary (right).
}
\end{figure}

Let us consider random walks on a planar lattice.  For those walks
returning to their starting point, we construct the outer boundary using
the following three-step algorithm.
\begin{itemize}
\item[(1)] Find the bottom vertex $v_0$ of the random walk.  This is
the first vertex in an increasing lexicographic ordering of the walk
vertices by their coordinates; first in the $x$-direction and then in
the $y$-direction.  Define the direction vector $e_{v_0}=-e_y$, where
$e_y$ is the unit vector in the $y$-direction. Set $i=0$.
\item[(2)] For the vertex $v_i$, consider the set $F_i$ of direction
vectors pointing from $v_i$ to those random walk vertices which are
also lattice nearest neighbours of $v_i$. Choose from $F_i$ the
direction vector $f_i$ of maximal angle $\alpha(e_{v_i},f_i)\in
[0,2\pi)$.  Set $v_{i+1}=v_i+f_i$ and $e_{v_{i+1}}=-f_i$.
\item[(3)] If $v_{i+1}\neq v_0$, then repeat step~(2), with $i$
replaced by $i+1$. Otherwise, set $N=i$ and stop.
\end{itemize}
The vertices $v_i$, subsequently obtained from the algorithm, yield a
closed walk $(v_0, v_1, \ldots, v_N)$, called the {\it outer boundary}
of the random walk.  We define its perimeter to be $N+1$. The walk travels
along the random walk in clockwise direction.  A typical square lattice
example is shown in Fig.~\ref{fig:snapshot}. 
Note that the outer boundary is not a self-avoiding polygon, since
it has cut vertices.  The simulations suggest however that such vertices occur
with zero density in the limit of infinite length of the outer boundary.

We considered the cases of the square lattice and the triangular
lattice.  The triangular lattice case was implemented on the computer using a
square lattice with additional edges along the positive diagonal of
each square.  Due to piecewise linearity, the area $A$ of a loop $(v_0, v_1,
\ldots, v_N)$ is, on both lattices, given by
\begin{displaymath}
A = \sum_{i=0}^N (v_{i+1,x}-v_{i,x})\frac{v_{i+1,y}+v_{i,y}}{2},
\end{displaymath}
where $v_{i,x}$ and $v_{i,y}$ denote the $x$-coordinate and $y$-coordinate of
the vertex $v_i$, and $v_{N+1}=v_0$.  It can thus be determined ``on
the fly'' together with the construction of the outer boundary.

We generated random walks up to length $10^5$ on the square and
triangular lattices, using the random number generator {\tt ran2.c}
\cite{PTVF92}. For each lattice, we generated $10^8$ random walks.
The size of the array was chosen such that the random walks did not
reach the array boundaries. In our situation, an array of size $5000 \times
5000$ was sufficient.  For those random walks returning to their start
point, we constructed the outer boundary and computed the values of
the first ten area moments for perimeter up to $10^3$, i.e., we
sampled $\mathbb E[Y^k_m]$ for $m<10^3$ and $k\le10$.  Note that only even
values of $m$ occur for the square lattice.  For the data analysis, we
rejected all values of $m$ with less than $10^4$ samples.  Furthermore, we
chose $m\ge50$ in order to reduce finite size effects.  This led to $50\le
m\le512$ for the square lattice and to $50\le m\le401$ for the triangular
lattice.  The total number of loops used in the data analysis was
$8.6\times 10^6$ for the square lattice and $11.6\times 10^6$ for the
triangular lattice.

We first analysed the predicted power-law behaviour of the area moments
(\ref{form:val}) 
\begin{equation}\label{form:Easym}
\mathbb E[Y_m^k] \sim A_k m^{\alpha_k} \qquad (m\to\infty),
\end{equation}
with $\alpha_k=3k/2$. Estimates of $\alpha_k$ were obtained by a least 
squares fit of the data to the assumed asymptotic form
\begin{equation}\label{form:fitting}
\log (\mathbb E[Y_{m}^k]) \sim \log A_k + \alpha_k \log m + 
\sum_{l=1}^N \frac{B_{k,l}}{m^{\beta_{k,l}}} \qquad (m\to\infty).
\end{equation}
Since the exponents of the corrections to the asymptotic behaviour of 
(\ref{form:val}) are not known, we chose $\beta_{k,l}=l$. This corresponds 
to analytic correction terms in (\ref{form:val}).  The results of a least 
squares fit with $N=3$ are shown in Table~\ref{tab:expo}.  The numbers in 
brackets give the statistical error of the fit in the last two digits.  
We found that differences in the estimates of $\alpha_k$ due to different
choices of the exponents $\beta_{k,l}$ are an order in magnitude smaller 
than the statistical errors. This reflects the observation that the 
statistical fluctuations of the data points dominate effects due to different 
corrections to the asymptotic behaviour (\ref{form:Easym}).  The obtained 
values are consistent with the prediction $\alpha_k=3k/2$, which we 
employ in the following.
\begin{table}
\begin{center}
\begin{tabular}{|c||c|c|c|c|c|}\hline
$k$ & 1 & 2 & 3 & 4 & 5\\ \hline
Square $\alpha_k$& 1.522(41) & 3.043(85) & 4.56(14) & 6.08(20) &
7.60(27)\\
Triangular $\alpha_k$& 1.508(81) & 3.02(16) & 4.53(24) & 6.04(34) &
7.56(45) \\ \hline
\end{tabular}
\end{center}
\begin{center}
\begin{tabular}{|c||c|c|c|c|c|}\hline
$k$ & 6 & 7 & 8 & 9 & 10\\ \hline
Square $\alpha_k$ & 9.12(36) & 10.63(48) & 12.14(62) & 13.65(80) &
15.1(10)\\
Triangular $\alpha_k$ & 9.08(58) & 10.60(73) & 12.11(90) & 13.6(11) &
15.2(14)\\ \hline
\end{tabular}
\end{center}
\caption{\label{tab:expo} Estimates of the exponents $\alpha_k$ of the
area moments $\mathbb E[Y_m^k]$ on the square and triangular lattices.  The
numbers in brackets denote the statistical error of the fit in the last two digits.}
\end{table} 

\begin{figure}[htb]
\begin{center}
\setlength{\unitlength}{1cm}
\begin{picture}(0,0)
\put(12,0){$1/m$}
\put(1.2,4.7){$\mathbb E[Y_m^2]/\mathbb E[Y_m]^2$}
\end{picture}
\begin{minipage}[b]{0.95\textwidth}
\center{\epsfig{file=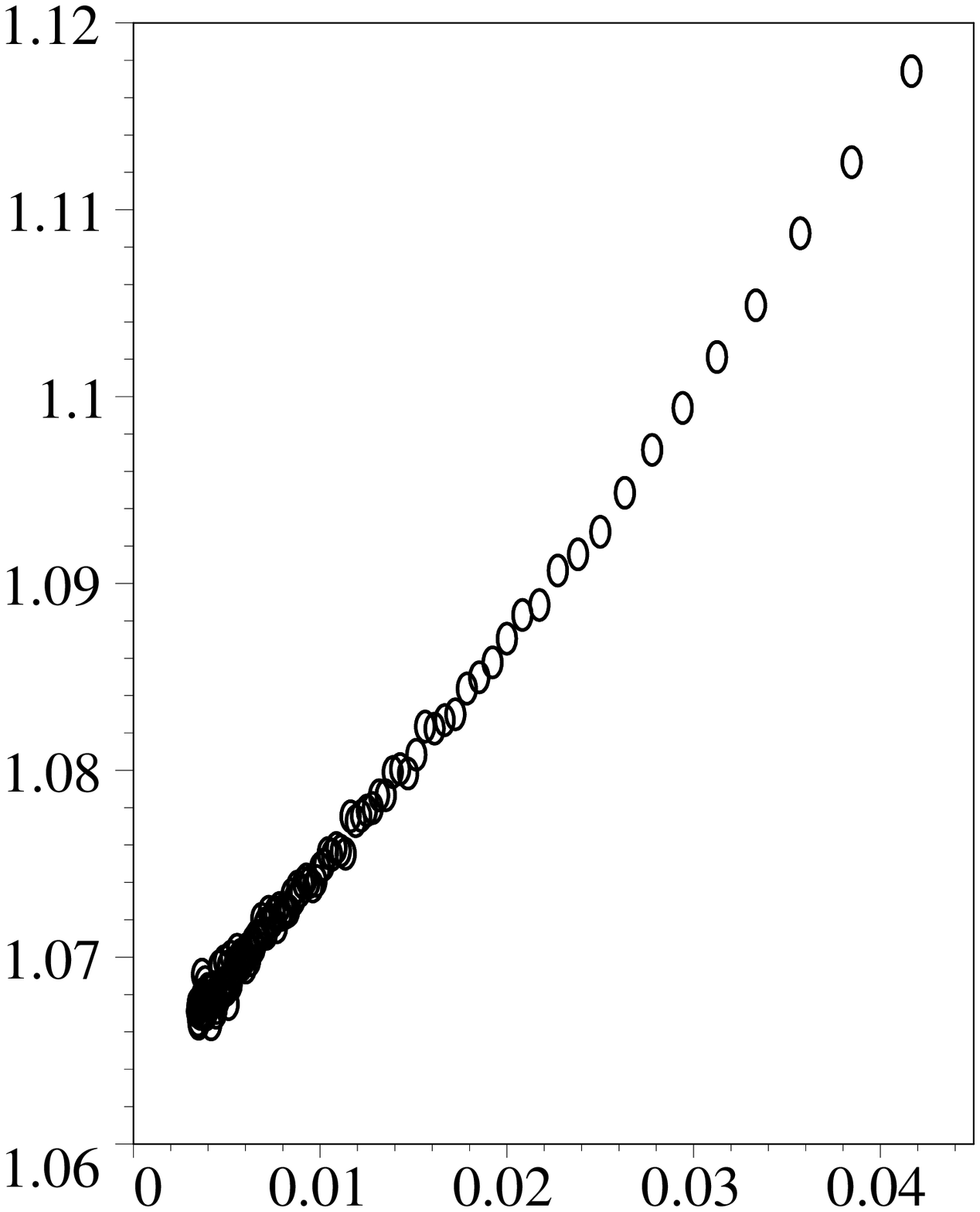,width=8cm, height=5cm}}
\end{minipage}
\end{center}
\caption{\label{fig:D2data}
\small MC estimates of $\mathbb E[Y_m^2]/\mathbb E[Y_m]^2$ on the
square lattice, plotted against $1/m$.}
\end{figure}

We then tested our prediction for the area moment amplitudes.  It
follows from (\ref{form:val}) that the ratios $A_k/A_1^k$ are
independent of the constant $C$.  In particular, we get
\begin{equation}
A_{2k}/A_1^{2k} = \frac{2^{6k-2}}{\pi^k}\frac{(3k-2)!}{(6k-3)!}\Omega_{2k}, \qquad
A_{2k+1}/A_1^{2k+1} = \frac{2}{\pi^k}\frac{1}{(3k)!}\Omega_{2k+1},
\end{equation}
where the numbers $\Omega_k$ are the Airy constants defined in
(\ref{form:om}).  We extrapolated the numbers $A_k/A_1^k$ by a least
squares fit to $\mathbb E[Y_m^k]/\mathbb E[Y_m]^k \sim
A_k/A_1^k+c_k/m+d_k/m^2$ as $m\to\infty$.
A typical data plot is shown in Figure \ref{fig:D2data}.

The corresponding results for the first ten amplitude ratios are
given in Table \ref{tab:data}, together with their predicted values.
The numbers in brackets give the statistical error of the fit in 
the last two digits.
\begin{table}
\begin{center}
\begin{tabular}{|c||c|c|c|c|}
\hline
Amplitude &  Airy distribution & Square MC & Triangular MC\\
\hline
$A_2/A_1^2$ &
1.06103296 &
1.0626(18) &
1.0620(23) \\
$A_3/A_1^3$ & 
1.19366208 &
1.1980(52) &
1.1951(61) \\
$A_4/A_1^4$ & 
1.42171312 &
1.430(13) &
1.422(15) \\
$A_5/A_1^5$ & 
1.78895216 &
1.801(28) &
1.783(31) \\
$A_6/A_1^6$ &
2.37205758 &
2.387(59) &
2.350(61) \\
$A_7/A_1^7$ &
3.30494239 &
3.32(12) &
3.24(12) \\
$A_8/A_1^8$ &
4.82446133 &
4.83(25) &
4.68(24) \\
$A_9/A_1^9$ & 
7.35752352 &
7.32(52) &
7.02(48) \\
$A_{10}/A_1^{10}$ &
11.6901282 &
11.5(11) &
10.93(98) \\
\hline
\end{tabular}
\end{center}
\caption{\label{tab:data} Comparison of amplitude ratios arising from
the Airy distribution against MC estimates for the square and
triangular lattice. The numbers in brackets denote the statistical error
of the fit in the last two digits.}
\end{table} 

Finally, we estimated the constant $C$ in (\ref{form:val}) using the
first area moment $\mathbb E[Y_m]\sim C \sqrt \pi\, m^{3/2}$ as $m\to\infty$.
A least squares fit to $\mathbb E[Y_m]/m^{3/2}$ with a polynomial of degree 4
in $1/m$ yields $C_{sq}=0.0910(11)$ for the square lattice and
$C_{tr}=0.1262(29)$ for the triangular lattice.

\section{Conclusion}

Our numerical analysis supports the conjecture that the area of the outer 
boundary of planar random loops is Airy distributed, on both the square 
and triangular lattices.  Since we expect this result to be universal, 
it would be interesting to check the appearance of the Airy distribution 
for boundaries of random walks on, for example, non-periodic tilings such 
as the Penrose tiling or the Ammann-Beenker tiling. Also, the question arises 
whether the Airy distribution can be found in other models of cluster 
boundaries with fractal dimension $4/3$.  The outstanding question is whether 
our conjecture can be proved, possibly by using quantum gravity methods 
\cite{Dup98} or the probabilistic approach \cite{LSW01}.

\section*{Acknowledgements}

The author thanks John Cardy, who pointed the author's attention to the
problem.  Financial support by the German Research Council (DFG) is gratefully
acknowledged.

\section*{Appendix: Scaling functions and limit distributions}

We explain how limit distributions are derived from the scaling function of a
combinatorial model.  We show, in particular, how the Airy distribution
is derived from the scaling function of polygon models \cite{J00} such as 
staircase polygons or self-avoiding polygons.  This relates the scaling 
function viewpoint employed in \cite{RGJ01,RJG03,J03} to the present approach. 
In particular, it motivates formula (\ref{form:val}).

For a polygon model on a given lattice, denote the number of polygons of 
(half-) perimeter $m$ and area $n$ by $p_{m,n}$.  The perimeter and area 
generating function of the model is given by $G(x,q)=\sum_{m,n} p_{m,n}x^m q^n$.  
The function $G(x,1)=g_0(x)$ is called the perimeter generating function.  
Let $x_c$ denote the radius of convergence of the perimeter generating 
function.  Typically \cite{R02}, the factorial area moment generating 
functions $g_k(x)$ have the asymptotic behaviour
\begin{equation}\label{form:gk}
g_k(x) =
\frac{(-1)^k}{k!}\left.\frac{\mbox{d}^k}{\mbox{d}q^k}G(x,q)\right|_{q=1}
\sim \frac{f_k}{(x_c-x)^{\gamma_k}} \qquad (x\to x_c^-),
\end{equation}
with critical amplitudes $f_k$ and critical exponents $\gamma_k$, where 
$\gamma_{k+1}>\gamma_k$.  See \cite[Ch.~7]{J00} for a proof that the 
functions $g_k(x)$ exist and are finite for polygon models.  It is 
convenient to incorporate the numbers $f_k$ and $\gamma_k$  into 
the function \cite{RGJ01,R02}
\begin{equation}
F(s)=\sum_{k=0}^\infty \frac{f_k}{s^{\gamma_k}}.
\end{equation}
Typically, the function $F(s)$ describes scaling behaviour of the perimeter
and area generating function about $(x,q)=(x_c,1)$,
\begin{equation}\label{form:scal}
G^{(sing)}(x,q) \sim (1-q)^\theta F\left( \frac{x_c-x}{(1-q)^\phi}\right) 
\qquad (x,q)\to (x_c^-,1^-),
\end{equation}
with critical exponents $\theta$ and $\phi$.  The function
$F(s)$ is also called the scaling function.  The form (\ref{form:scal})
is consistent with (\ref{form:gk}) if the exponents $\gamma_k$ are of the form
$\gamma_k=(k-\theta)/\phi$. One should note that (\ref{form:scal}) has been
proved only for a few examples including staircase polygons \cite{P95}.  For 
typical polygon models such as staircase polygons%
\footnote{
More generally, for polygon models satisfying a $q$-algebraic functional
equation with a square-root singularity as dominant singularity of their
perimeter generating function \cite{R02}.}
and rooted self-avoiding polygons \cite{RGJ01}, the
scaling function has the particular form \cite[Eqn.~3.9]{R02}
\begin{equation}\label{form:log2Ai}
F(s)=-4f_1 \frac{\rm d}{{\rm d}s}\log \mbox{Ai}\left( \left(
    \frac{f_0}{4f_1}\right)^{2/3}s\right),
\end{equation}
with critical exponents $\theta=1/3$ and $\phi=2/3$.

We now turn to a probabilistic description, see also \cite{D99}.  Selecting
polygons uniformly among the set of all polygons with perimeter $m$ defines a
discrete random variable $X_m$ with probability distribution
\begin{equation}
\mathbb P(X_m=n)=\frac{p_{m,n}}{\sum_n p_{m,n}},
\end{equation}  
whose moments are given by
\begin{equation}\label{form:exp}
\mathbb E[X_m^k] = \frac{\sum_n n^k p_{m,n}}{\sum_n p_{m,n}}.
\end{equation}
The numerator in (\ref{form:exp}) is related to the coefficients of the
factorial moment generating functions (\ref{form:gk}) via
\begin{equation}\label{form:asympt}
\frac{(-1)^k}{k!}\sum_n n^k p_{m,n} \sim
\frac{(-1)^k}{k!}\sum_n (n)_k \, p_{m,n} = [x^m]g_k(x)\sim
\frac{f_k}{x_c^{\gamma_k}\Gamma(\gamma_k)}x_c^{-m} m^{\gamma_k-1}
\qquad (m\to\infty),
\end{equation}
where $(n)_k=n(n-1)\cdots(n-k+1)$, and $\Gamma(z)$ is the Gamma function.  This
relation can be used to characterise the leading asymptotic behaviour
of the moments $\mathbb E[X_m^k]$ of the underlying distribution.  Let us 
introduce distribution coefficients $\Phi_k$ defined by
$f_k=-\frac{(-1)^{k}}{k!} 2^k \Phi_k f_1^k f_0^{1-k}$.  We obtain
\begin{equation}\label{form:asympt3}
\mathbb E[X_m^k] \sim -\frac{\Gamma(\gamma_0)}{\Gamma(\gamma_k)}
\left( \left( \frac{m}{x_c}\right)^{1/\phi} \frac{2f_1}{f_0}\right)^k 
\Phi_k \qquad (m\to\infty).
\end{equation}
Now introduce a normalised random variable ${\widetilde X}_m$ defined by
\begin{equation}
{\widetilde X}_m = \frac{X_m}{\left( \frac{m}{x_c}\right)^{1/\phi}
\frac{2f_1}{f_0}}.
\end{equation}
The moments $\mathbb E[{\widetilde X}_m^k]$  of the normalised random variable
${\widetilde X}_m$ are then asymptotically given by
\begin{equation}\label{form:asympt2}
\mathbb E[{\widetilde X}_m^k] \sim -\frac{\Gamma(\gamma_0)}{\Gamma(\gamma_k)} 
\Phi_k \qquad (m\to\infty).
\end{equation} 
Note that (\ref{form:asympt2}) is independent of $x_c$ and asymptotically
constant. In conjunction with (\ref{form:logAi}), it follows that the 
particular scaling function (\ref{form:log2Ai}) leads to the Airy distribution 
(\ref{form:Airydistr}) in (\ref{form:asympt2}).  Formula (\ref{form:val}) is 
chosen in analogy with (\ref{form:asympt3}).

\end{document}